\documentstyle[preprint,
prb,aps]{revtex}

\newcommand{\cer}{\mathrm{Ce_{2}Rh_{3}Al_{9}}}
\newcommand{\TK}{T_{\mathrm{K}}}
\newcommand{\EF}{E\mathrm{_{F}}}
\newcommand{\Uff}{\mathit{U_{ff}}}

\begin{document}

\draft

%\wideabs{

\title{Ce-$L_3$-XAS study of the temperature dependence of the $4f$ 
occupancy in the Kondo system Ce$_{2}$Rh$_{3}$Al$_{9}$}

\author{C. Grazioli, Z. Hu, M. S. Golden, and J. Fink}
\address{Institute for Solid State and Materials Research Dresden, P.O. Box
270016, D-01171 Dresden, Germany}

\author{K. Attenkofer}
\address{HASYLAB at DESY, Notkestrasse 85, D-22603 Hamburg, Germany}

\author{O. Trovarelli and C. Geibel}
\address{Max Planck Institute for Chemical Physics of Solids, Dresden}
\maketitle

\begin{abstract}

We have used temperature dependent x-ray absorption at the Ce-$L_3$ edge 
to investigate the recently discovered Kondo compound Ce$_{2}$Rh$_{3}$Al$_{9}$.
The systematic changes of the spectral lineshape with decreasing temperature are analyzed and found
to be related to a change in the $4f$ occupation number, $n_f$, as the system undergoes a transition 
into a Kondo state.
The temperature dependence of $n_f$ indicates a characteristic temperature of 150~K, which is clearly related
with the high temperature anomaly observed in the magnetic susceptibility of the same system.
The further anomaly observed in the resistivity of this system at low temperature (ca. 20~K) has no effect on
$n_f$ and is thus not of Kondo origin. 


\end{abstract}
%}

%\section{Introduction}

The most successful description of Kondo systems is that of the Anderson impurity model (AIM).
In this model the energy of the $4f$ levels and their hybridization with the 
conduction band states are taken into account to explain the low temperature 
behavior of the resistivity and the magnetic susceptibility. 
The coupling between $4f$ levels and band states leads to the formation of a 
non-magnetic, singlet ground state on a temperature scale characterized by $\TK$, 
the Kondo temperature. The theory also provides two predictions: 
firstly the formation of the singlet state leads to the occurance 
of the Abrikosov-Suhl peak at $\EF$ in the density of states
(the Kondo resonance). Secondly, the $4f$ occupation, $n_f$,
should show a characteristic temperature dependence as a function of 
$T/\TK$ \cite{Bickers1987}. 

RE-$L_3$ x-ray absorption (XAS) is very sensitive to the distribution of the valence electrons\cite{Malterre1991,Kaindl1988,Hu95,RoehlerRev} 
and has therefore often been used 
to study the change of $n_f$ with temperature. 
The advantages of this technique are: 
1) the clear signature of the $4f^0$ configuration 
on the high-energy side of the main line; 
2) the negligible role played by multiplet effects; 
3) the absence of surface related effects. 

Recently we have applied this technique to the classical 
Kondo system CeSi$_2$\cite{Grazioli2001}, and found that the extracted $n_f$ shows 
a characteristic temperature dependence as expected by the AIM, 
whereby the strong change in $n_f$ is centered at a value of $\TK$ which is 
fully compatible with the Kondo temperature as determined by thermodynamic measurements. 

\par
In this work we report Ce-$L_3$ XAS results from the new Kondo system $\cer$, 
which crystallizes in a complicated orthorhombic (Cmcm) 
Y$_2$Co$_3$Ga$_9$-structure, with only one Ce site\cite{Buschinger1997}. 
Neutron diffraction experiments of $\cer$ show no magnetic ordering down to 
1.8~K\cite{Hiess1999}. The susceptibility data present a maximum at 150~K, while the 
transport data show the occurrence of two clear anomalies at 20~K 
and 150~K\cite{Buschinger1997,Buschinger1998}. 
Although a satisfactory explanation of the behavior of the $4f$ Ce states in this 
compound is still to be found, this unconventional situation might invoke 
the necessity of a description in terms of two different energy scales for the 
same $4f$ electrons. An alternative explanation in analogy 
with heavy fermion systems which have Kondo temperatures
of about 20~K and crystal field excitations at around 150~K 
seems unlikely on the basis of magnetic susceptibility considerations\cite{Buschinger1998}.

%\section{Experimental}

\par
Powder samples were synthesized by repeated argon arc melting of the constituent
elements, followed by annealing in vacuum at high $T$ for 120 h.
X-ray powder diffraction measurements taken immediately after melting showed that the samples were
phase pure. 
The Ce-$L_3$ XAS experiments were performed at the 
A1 beamline in HASYLAB in the transmission geometry with 
the help of three ionization chambers in series for the sample 
and a reference. The overall resolution at the Ce-$L_3$ edge 
(5720~eV) was estimated to be about 1.5 eV. 
The measurements were recorded within a few days of the sample preparation, and concurrently
we detected no appreciable change due to sample oxidation. 
We also prepared homogeneous absorbers of polycrystalline 
CeF$_3$ and CeO$_2$ as trivalent and tetravalent Ce reference 
compounds, respectively. 

%\section{Results}

\par

Fig.~\ref{fig1} shows the Ce-$L_3$ spectrum of $\cer$ 
together with those of CeF$_3$, 
CeO$_2$ and CeSi$_2$, measured at room temperature. 
The spectrum of $\cer$ shows clearly the characteristic signature 
of the $4f^0$ final state in the form of the shoulder seen at 5735 eV, 
which is well known from previous studies of mixed valent Ce compounds. 
The main line is due to the $4f^1$ final state, mixed with 
$4f^2$. Since the $4f$ states are characterized by a stong 
on-site $4f$ electron-electron repulsion $\Uff$ (about 10 eV), 
the contribution of $4f^2$ is small and the contribution from higher $4f$
occupancies is negligible. 
$\cer$ shows a more pronounced high-energy shoulder than CeSi$_2$, indicating
that the $4f^0$ configuration plays a more important role in the room temperature ground state 
of the former. 

\par

Fig.~\ref{fig2} shows the Ce-$L_3$ XAS spectra of $\cer$ taken at 
temperatures between 6 and 240~K. 
From Fig.~\ref{fig2} one can clearly see a change of the relative intensity ratio 
of the shoulder structure compared to the main peak at 5726 eV as the temperature is
altered.
As mentioned above, although the main peak and shoulder are dominated by the $4f^1$ 
and $4f^0$ final state configurations, respectively, there is also a non-zero $4f^2$ admixture.
Therefore the spectral weight of the high-energy shoulder alone 
does not represent an accurate measure of $n_f$ in the initial state. 
In order to take this into account, we analyse the data using a simplified version of the Anderson
single-impurity Hamiltonian\cite{Imer1987}.
In this analysis, each final state itself is represented by a Lorentzian function with an arctan-like 
step function describing the edge jump (for details see Refs.~\onlinecite{Hu95,Grazioli2001}). 
To reduce the number of free parameters to a minimum, the energy separation between the Lorentzian 
and the arctan step function, as well as the ratio of the Lorentzian intensity 
to the arctan step height were kept equal for all final state components. 
Finally, the theory data were broadened using a Gaussian function of width 1.5~eV to account for the
finite experimental energy resolution.

\par
In Fig.~\ref{fig3} we plot the temperature dependence 
of the $4f$ occupation number, $n_f$, deduced using this procedure. 
We obtain a zero temperature extrapolation $n_f$($T$=0) of 0.806, which is smaller 
than the 0.9 obtained for CeSi$_2$\cite{Grazioli2001}, which is in keeping with the 
qualitative comparison the spectral profiles as discussed above. 
As the temperature increases, one can clearly see
an increase in $n_f$. In the XAS spectra of Fig. 2, this is evident as a reduction of the $4f^0$- related
shoulder for $T>\TK$. 
This increase in $n_f$ is centered at $\sim$~150~K, which is consistent with the anomaly observed in the
susceptibility data. This result indicates that the anomaly at 150~K is not due to a crystal field effect, but
rather is related to a change in the $4f$ occupancy. The onset of this change in $n_f$ is at about 70~K, and no
variation in $n_f$ could be detected for $T \ll 60$~K. Thus, the low temperature anomaly observed in this system in
transport measurements is not connected to a change in the $4f$ occupancy. This implies that of the two
energy scales shown by the system in transport and susceptibility measurements, only one (the higher
energy one) is related to Kondo behavior.

%\section{Summary}

In summary, we have used bulk sensitive Ce-$L_3$ XAS to measure a powdered single crystal of $\cer$ and 
we have shown that the Ce $4f$ occupancy, $n_f$, shows a temperature dependent behavior which is in good agreement 
with expectations from the single Anderson impurity model with a Kondo temperature $\TK \sim$~150~K.
This is the same temperature at which anomalies are seen in both susceptibility and transport measurements, thus 
putting beyond doubt the Kondo origin of these anomalies.
However, no change in $n_f$ is found at temperatures near to the low temperature anomaly observed in transport 
data, which calls for further experimental and theoretical work aimed at understanding this intriguing observation.

%\section{Acknowledgments}

This work was supported by the Deutsche For\-schungs\-ge\-mein\-schaft 
(DFG) within the SFB 463 \it "Sel\-ten\-erd-\"Uber\-gangs\-me\-tall\-ver\-bin\-dungen: Struk\-tur, 
Mag\-ne\-tis\-mus und Trans\-port"\rm.
C.G.~is grateful to Dr.~P.~Thalmeier for valuable discussions.

\begin{figure}
\caption{
Ce-$L_3$ x-ray absorption spectra of $\cer$ recorded room temperature together with those
of CeSi$_2$ (from Ref.~\protect\onlinecite{Grazioli2001}) and the Ce(III) and Ce(IV) standards
CeF$_3$ and CeO$_2$. 
}
\label{fig1} 
\end{figure}

\begin{figure}
\caption{Ce-$L_3$ x-ray absorption spectra of $\cer$ at various temperatures. 
A fit of the main line (labelled $|f^1\rangle$) and the edge-jump is also shown as a solid line. 
Comparison between the data and the fit clearly shows the change in the intensity
of the $\underline{2p}\;4f^0 5d^*$-related final state feature. 
}
\label{fig2} 
\end{figure}

\begin{figure}
\caption{
Ce $4f$ occupancy, $n_f$, extracted from the Ce-$L_3$ x-ray absorption data
of Fig.~2 versus $T$. The dotted vertical line indicates the center of the 
strong temperature dependence of $n_f$ at ca. 150K.
}
\label{fig3} 
\end{figure}

\end{document}